\documentstyle[12pt,epsfig]{article}
\def\del{{\partial}}

\newfont{\Bbb}{msbm10 scaled 1200}     %instead of eusb10
\newcommand{\mathbb}[1]{\mbox{\Bbb #1}}

\def\Poincare{{Poincar\'e }}

\def\lbldef#1#2{\expandafter\gdef\csname #1\endcsname {#2}}
\def\eqn#1#2{\lbldef{#1}{(\ref{#1})}%
\begin{equation} #2 \label{#1} \end{equation}}

\def\href#1#2{#2}

\newcommand{\beq}{\begin{equation}}
\newcommand{\eeq}{\end{equation}}
\newcommand{\ber}{\begin{eqnarray}}
\newcommand{\eer}{\end{eqnarray}}

\newcommand{\beqar}{\begin{eqnarray}}

\newcommand{\eeqar}{\end{eqnarray}}

\begin{document}
\baselineskip=15.5pt
\pagestyle{plain}
\setcounter{page}{1}
%\renewcommand{\thefootnote}{\fnsymbol{footnote}}
%--------+---------+---------+---------+---------+---------+---------+
%Title page
\begin{titlepage}

\leftline{\tt hep-th/0005183}

\vskip -.8cm

\rightline{\small{\tt CALT-68-2266}}
\rightline{\small{\tt CITUSC/00-021}} 
\rightline{\small{\tt HUTP-00/A009}}
\rightline{\small{\tt  UCB-PTH-00/10}}

\begin{center}

\vskip 1.7 cm

{\LARGE {Strings in $AdS_3$ and the $SL(2,R)$ WZW Model.}}
\vskip .5cm 
{\LARGE{Part 2: Euclidean Black Hole}}

\vskip 1.5cm
{\large 
Juan Maldacena$^{*}$,
Hirosi Ooguri$^{\dagger}$\footnote{On leave of absence
from the University of California, Berkeley.}, and
John Son$^{*}$}

\vskip 1.2cm

${}^*$ Lyman Laboratory of Physics,
Harvard University, Cambridge, MA  02138, USA

{\tt malda, json@pauli.harvard.edu}

\medskip
\vskip .5cm

${}^\dagger$Caltech-USC Center for Theoretical Physics, Mail Code 452-48\\
California Institute of Technology, 
Pasadena, CA 91125, USA 

{\tt ooguri@theory.caltech.edu}

\vskip 1.7cm

{\bf Abstract}
\end{center}

\noindent

We consider the one-loop partition function for 
Euclidean BTZ black hole backgrounds or equivalently 
thermal $AdS_3$ backgrounds which are quotients of 
$H_3$ (Euclidean $AdS_3$). The one-loop partition 
function is  modular invariant and we can read off 
the spectrum which is consistent to that found in 
hep-th/0001053. We see long strings and discrete 
states in agreement with the expectations. 

\end{titlepage}

\newpage

%--------+---------+---------+---------+---------+---------+---------+
%Body

\section{Introduction}
\label{intro}

In this paper we continue the investigation started in 
\cite{Maldacena:2000hw}
of the $SL(2,R)$ WZW model describing string theory on 
$AdS_3 \times {\cal M}$. For other work on this model,  
see \cite{list}. 
Our motivation is to understand string theories in 
curved spacetimes where the metric component $g_{00}$
is non-trivial, of which $AdS_3$ is the simplest 
example.  Moreover, it is possible to construct black hole 
solutions as quotients of $AdS_3$ \cite{btz}, so understanding 
string theory on $AdS_3$ would lead to an understanding of 
strings moving near black hole horizons. 

In \cite{Maldacena:2000hw} the spectrum
 of $SL(2,R)$ WZW model was studied, using 
spectral flow to generate new representations from the standard 
ones.  These new representations include states corresponding
to long strings \cite{Maldacena:1998uz,Seiberg:1999xz},
 with a continuous energy spectrum, as well as
discrete states.  The existence of spectral flow as a symmetry of the 
theory was argued on the basis of classical and semi-classical analysis.
Further support was given by the fact that the seemingly arbitrary 
upper bound on the mass of string states in $AdS_3$ was removed,
thus recovering the infinite tower of masses one expects from string 
theory.  We would like to verify these results by an explicit 
calculation of the one-loop partition function. 
As shown in  \cite{malda}, the Euclidean black hole background is 
equivalent to the thermal $AdS_3$ background. 
So we will consider string theory on $AdS_3$ at a finite temperature, 
which is described by strings moving on a Euclidean $AdS_3$ background
with the Euclidean time identified. 
The calculation of the partition function for this geometry is a 
minor variation on the calculation of
Gawedzki in \cite{Gawedzki:1991yu}. From 
this we can read off the spectrum of the theory
in Lorentzian signature by interpreting the result as the
free energy of a gas of strings.

This paper is organized as follows.
In section 2 we review the spectrum found in \cite{Maldacena:2000hw}.
In section 3 we compute the one-loop partition function on thermal 
$AdS_3$. In section 4 we read off the spectrum from the one-loop
calculation. First we present a qualitative analysis, which is then
followed by a precise calculation. We explain how the different parts
of the spectrum arise from this calculation. We further show how
the one-loop result contains information about the 
$SL(2,R)$ and Liouville reflection amplitudes.

\section{ The spectrum }

We begin by briefly summarizing the results of \cite{Maldacena:2000hw},
 where  
a concrete  proposal for the spectrum of $AdS_3$ string theory 
was made. We consider a critical 
bosonic string theory on $AdS_3 \times {\cal M}$. 
 The Hilbert space of
the $SL(2,R)$ WZW model is generated by 
the action of the   left-moving and  
right-moving current algebra 
$\widehat{SL(2,R)}_L \times \widehat{SL(2,R)}_R$, and
all the states form  representations of this algebra.
  The simplest  representations are built by first 
choosing representations for the zero modes, then regarding them as the 
primary states annihilated by $J^{3,\pm}_{n > 0}$.  
The raising operators $J^{3,\pm}_{n < 0}$ are then used to generate 
the representations of the current algebra.  From harmonic analysis,  
i.e.\ quantum mechanical limit, it is known that the left-right 
symmetric combinations 
${\cal C}^{\alpha}_{j=1/2 + is} \times {\cal C}^{\alpha}_{j=1/2 + is}$ 
and ${\cal D} ^{\pm}_{j >1/2} \times {\cal D}^{\pm}_{j >1/2}$ form a 
complete basis in ${\cal L}^2(AdS_3)$, where 
${\cal C}^{\alpha}_{j=1/2 + is}$ is the principal continuous representation 
and ${\cal D}^{\pm}_{j >1/2}$  the principal discrete representation of 
$SL(2,R)$.  These representations are unitary, but the resulting current 
algebra representations 
$\widehat{{\cal C}}^{\alpha}_{j=1/2 + is} \times
\widehat{{\cal C}}^{\alpha}_{j=1/2 + is}$ and 
$\widehat{{\cal D}} ^{\pm}_{j >1/2} \times 
\widehat{{\cal D}}^{\pm}_{j >1/2}$, 
constructed as explained above, are not.  This is not a surprise, for even 
in flat Minkowski space it is not until one imposes the Virasoro constraints
\eqn{Vir}{
(L_n+{\cal L}_n - \delta _{n,0})|{\rm physical} \rangle = 0, \ \ n \geq 0
}
that a unitary spectrum is obtained.  Here ${\cal L}_n$ is the
 Virasoro generator for the internal conformal field 
theory corresponding to ${\cal M}$.  
The proposal of \cite{Maldacena:2000hw} is that 
one should consider not just these representations 
 but also those obtained by the spectral flow
\ber \label{spect}
J^3_n &\rightarrow & \tilde{J}^3_n = J^3_n - {k \over 2}w \delta _{n,0} 
\nonumber \\
J^+_n &\rightarrow & \tilde{J}^+_n = J^+_{n+w}   \\
J^-_n &\rightarrow & \tilde{J}^-_n = J^-_{n-w}. \nonumber
\eer
The Virasoro generators, given by the Sugawara form, then become 
$\tilde{L}_n = L_n + wJ^3_n -{k \over 4}w^2 \delta _{n,0}$.  Imposing on 
$\widehat{{\cal D}} ^{\pm}_{j >1/2} \times \widehat{{\cal D}}^{\pm}_{j >1/2}$ 
the condition \Vir \ with $\tilde{L}_n$ one 
finds that these states have a discrete energy spectrum 
\ber \label{edisc}
E & = & J^3_0 + {\bar J}^3_0 =q + \bar q + k w + 2 \tilde j  
\nonumber \\
& = & 1+q+\bar{q}+2w+\sqrt{1+4(k-2)\left(N_w+h-1-{1 \over 2}w(w+1)\right),}
\eer
here $N_w$ is defined to be the level of the current algebra after spectral
flow by amount $w$, $N_w = \tilde N - wq$, and $\tilde N$ is the level
before
spectral flow. The state with energy (\ref{edisc}) is obtained
from a lowest weight state by acting with the $SL(2,R)$ currents
$\prod {\tilde J}^\pm_{n\leq 0} | \tilde{j},\tilde{j}\rangle $,
with $q$ the net number
of $\pm$ signs in this expression. In other words, $q$ is the 
number of spacetime energy raising operators $J^+_a$ minus the 
number of spacetime energy lowering operators $J^-_a$ that we
have to apply to the lowest weight, lowest energy state $ |\tilde{j}, 
m = \tilde{j} \rangle$
to get to the state whose spacetime energy is (\ref{edisc}). 
$\bar q$ is the corresponding quantity for the generators $\bar J^\pm_a$. 
We also have a level matching condition of the form 
\eqn{levelms}{
 N_w + h = \bar N_w + \bar h }
which implies that the angular momentum in $AdS_3$, $\ell = 
J^3_0 - \bar J^3_0 = q -\bar q $, is an integer.
  We argued in \cite{Maldacena:2000hw} that 
$\tilde j$ is further restricted to the range 
\eqn{rangej}{
{1 \over 2}< \tilde j < {k-1 \over 2} \ , }
 which implies
\eqn{rangedisc}{
{k \over 4}w^2 + {1 \over 2}w < N_w + h -1+ {1\over 4(k-2)} 
< {k\over 4}(w+1)^2 - {1\over 2}(w+1).
}
A similar analysis on 
$\widehat{{\cal C}}^{\alpha}_{j=1/2 + is} 
\times \widehat{{\cal C}}^{\alpha}_{j=1/2 + is}$ 
yields a continuous spectrum 
\eqn{econt}{
E={k\over 2}w +{1 \over w}\left({ 2s^2+{1\over2}  \over k-2} + 
\tilde N +h + \tilde {\bar N} + \bar h - 2 \right),
}
where $s$ takes values over the real numbers and is interpreted as the 
momentum in radial direction for the long strings. 
These states satisfy the level matching condition 
\eqn{levelml}{
\tilde N + h = \tilde {\bar N} +  \bar h + w \times ({\rm integer}).
}
In the rest of the paper we will do an independent calculation 
which will reproduce this single string spectrum. 

\section{One-loop partition function}
\label{oneloop}

In this section we compute the worldsheet 
 one-loop partition function. 
 First we explain the relation between  various useful
 coordinate systems. Then we consider thermal $AdS_3 = H_3/Z$ and
 show how the identification of Euclidean 
time in the global coordinates translates into
 particular   boundary conditions 
for the target space fields.  The partition function is then 
calculated by an explicit evaluation of the functional 
integral following \cite{Gawedzki:1991yu}.

\subsection{Coordinates on $H_3$ and thermal $AdS_3$. }

The natural metric on $H_3$ is given by 
\eqn{poinc}{
ds^2= {k \over y^2} (dy^2+dwd\bar{w}),
}
which is the Euclidean continuation of the \Poincare metric on $AdS_3$.  
By the coordinate transformation
\ber
w      & = & \tanh\rho e^{t+i\theta}  \nonumber \\
\bar{w}& = & \tanh\rho  e^{t-i\theta}  \\
y      & = & {e^t \over \cosh \rho}   \nonumber 
\eer
we obtain the cylindrical coordinates on Euclidean $AdS_3$,
\eqn{cyl}{
{ds^2\over k} = \cosh^2 \rho  dt ^2 + d\rho^2 + 
\sinh^2 \rho  d\theta ^2.
}
For the purpose of calculating the partition function, however,
 it is convenient to use  coordinates in which the metric reads 
\cite{Gawedzki:1991yu}
\eqn{gawmetric}{
{ds^2\over k}= d\phi ^2 + (dv + v d\phi)(d\bar v +  \bar v
d\phi),
}
which corresponds to the parametrization of $H_3$ as 
\eqn{gawparam}{
g= \left[ \begin{array}{cc}
            e^{\phi}(1+|v|^2)    & v  \\
            \bar{v}              & e^{-\phi}
                  \end{array} \right].
}
The coordinate transformation from \cyl\ to \gawmetric\ is
\ber
v       & = & \sinh \rho e^{i \theta}       \nonumber  \\
\bar{v} & = & \sinh \rho e^{-i \theta}             \\
\phi    & = & t - \log \cosh \rho ~ .   \nonumber
\eer
Thermal $AdS_3$ is given by the identification
\eqn{tauiden}{
t + i \theta \sim t + i\theta + \hat{\beta} \ ,
}
where $\hat{\beta}$ represents the temperature $T$ 
and the imaginary chemical potential $i\mu$ for the angular
momentum, 
\eqn{temp}{
\hat{\beta} = \beta + i\mu\beta = {1 \over T} + i{\mu \over T} \ .
}
The corresponding identifications in the coordinates {\gawmetric} are
\ber \label{iden}
v       & \sim & ve^{i\mu\beta}            \nonumber \\
\bar{v} & \sim & \bar{v}e^{-i\mu\beta}       \\  
\phi    & \sim & \phi + \beta \ ,           \nonumber 
\eer
which is a consistent symmetry of the WZW action,
\eqn{act}{
S = {k \over \pi} \int d^2z \left( \del \phi \bar{\del} \phi + (\del \bar v + \del
\phi\bar{v})(\bar{\del} v  + \bar{\del} \phi v) \right) \ .
}

\subsection{Computation of the partition function on thermal $AdS_3$.}
 
In this subsection we compute the  partition function for string theory 
on thermal $AdS_3$. We consider a conformal field theory on a worldsheet 
torus with modular parameter $\tau$ ($z \sim z + 2\pi \sim z + 2 \pi \tau$). 
The two-dimensional conformal field theory on the worldsheet is the sum
of three parts: the conformal field theory on $H_3$, the internal conformal
field
theory on ${\cal M}$, and the $(b,c)$ ghosts. 
First we start with the computation of the partition function for the
conformal field theory describing the three dimensions of thermal $AdS_3$ 
and then we will multiply the result by the partition
function of the ghosts and the internal conformal field theory. 

Due to the identification (\ref{iden}), the string coordinates
now satisfy the following  boundary conditions  
\ber 
\phi(z + 2\pi)  &=& 
\phi( z) +\beta n,~~~\phi(z + 2\pi\tau) = 
\phi( z) + \beta m, \nonumber\\
v(z+ 2\pi) 
&=& v(z)e^{i n \mu\beta},~~~ v(z+2\pi\tau) =
v( z) e^{i m \mu\beta }.
\eer
The thermal circle is non-contractible and therefore we get two
integers $(n,m)$ characterizing topologically nontrivial embeddings of
the worldsheet in spacetime.  In order to implement these 
boundary conditions it is convenient to define new fields 
$ \hat{\phi} , \hat{v}$ such that they 
are periodic: 
\ber 
\phi & = & \hat{\phi} + \beta f_{n,m}(z,\bar{z})  \nonumber \\
v    & = & \hat{v} \exp(i \mu\beta f_{n,m}(z,\bar{z}) ),
\eer
with
\eqn{hnm}{
f_{n,m}(z,\bar{z}) = {i \over 4\pi \tau_2}\left[z(n\bar{\tau}-m)-\bar{z}(n\tau
-m)\right] \ .
}
When we substitute this into the action \act , we get
\eqn{act2}{
S = 
{k\beta^2 \over 4\pi \tau_2}|n\tau - m|^2 + {k \over \pi} \int 
d^2z \left( |\partial \hat{\phi}|^2 + \left| 
\left(\partial  + {1\over 2 \tau_2}
U_{n,m} + \partial \hat{\phi}\right) \hat{\bar v}\right|^2 \right),
}
where 
\eqn{unm}{
U_{n,m}(\tau) = {i\over 2\pi}(\beta-i\mu\beta)(n \bar{\tau} -m). 
}
We are interested in the functional integral
\eqn{path}{
{\cal Z}(\beta,\mu; \tau)= \int {\cal D} \phi {\cal D}  
v {\cal D} \bar v e^{-S}  \ .
}
This integral is evaluated as explained in \cite{Gawedzki:1991yu}. We can 
first do the integral over $\hat{v},\hat{\bar v}$ which is quadratic, 
giving the determinant
\eqn{detv}{
\det \left|\partial + {1\over 2 \tau_2}U_{n,m} + \partial \hat{\phi}
\right|^{-2}.
}
We calculate  the $\hat{\phi}$ dependence on the determinants
by realizing that we can view \detv\ as an inverse of two 
fermion determinants. We can then remove  $\hat \phi$ from the determinants
by  a chiral gauge transformation and using the formulas for chiral 
anomalies.  The result is
\eqn{poly}{
\det \left|\partial + {1\over 2 \tau_2}U_{n,m} + \partial \hat{\phi}
\right|^{-2} = e^{{2 \over \pi}\int d^2z 
\partial  \hat{\phi} \bar{\partial}
\hat{\phi} }\ \det \left|\partial + {1\over 2 \tau_2}U_{n,m}
 \right|^{-2}.
}
The remaining integral over $\hat{\phi}$ gives the usual result
for a free  boson, except that $k\rightarrow k -2$ due
 to \poly . The functional integral for the thermal $AdS_3$ partition 
function then gives 
\ber \label{partads}
&& {\cal Z}(\beta,\mu;\tau) \nonumber \\
&=& {\beta (k-2)^{{1\over 2}}\over 8\pi\sqrt{\tau_2} }  \sum_{n,m} 
{ e^{- k \beta^2 | m - n \tau|^2/ 4\pi\tau_2
+ 2\pi({\rm Im}  U_{n,m})^2 /\tau_2}  \over 
|\sin(\pi U_{n,m})|^2 |\prod_{r=1}^\infty 
(1-e^{2\pi i r\tau})
(1-e^{2\pi i r\tau+ 2\pi iU_{n,m}})(1-e^{2\pi i r \tau-2\pi iU_{n,m}})|^2 } 
\nonumber \\ 
&=& {\beta (k-2)^{{1\over 2}}\over 2\pi\sqrt{\tau_2}}   
 (q\bar{q})^{-{3 \over 24}}\sum_{n,m}  {  
 e^{- k \beta^2 | m - n \tau|^2/ 4\pi\tau_2+
2\pi({\rm Im} U_{n,m})^2 /\tau_2} 
 \over  |\vartheta_1( \tau, U_{n,m} )|^2 },
\eer
where $\vartheta_1$ is the elliptic theta function and $q=e^{ 2 \pi i \tau}$.  
The factor $\beta(k-2)^{{1\over 2}}$ comes from the length of the circle in
the $\phi$ direction. 
This partition function is explicitly modular invariant after summing over
$(n,m)$\footnote{In our previous paper,
there was a puzzle about the apparent lack of modular
invariance of the $SL(2,R)$ partition functions
with $J^3$ insertions (see Appendix B of \cite{Maldacena:2000hw}). 
Here we have found that, if we introduce the twist by considering 
the physical set-up of 
thermal $AdS_3$, the result (\ref{partads}) turns out to be
manifestly modular invariant. This resolves the puzzle raised in
\cite{Maldacena:2000hw}.}.

We also need to include the contribution of the $(b,c)$ ghosts and the 
internal CFT.  Partition function of the latter will be of the form 
\eqn{,}{
{\cal Z}_{\cal M} =  (q\bar{q})^{- {c_{int} \over 24}} \sum_{h, \bar{h}}
D(h, \bar{h}) q^h \bar{q}^{\bar{h}},
}
where $D(h, \bar{h})$ is the degeneracy at left-moving weight $h$ and
 right-moving weight $\bar{h}$, and $c_{int}$ the central charge of the 
internal CFT.  Modular invariance  requires that $h -\bar{h} \in
Z$,
 a fact which will be useful in the next section. 
 Vanishing of the total 
conformal anomaly gives
\eqn{ctotal}{
c_{SL(2,R)} + c_{int} = 26.
}
We can calculate now the total 
contribution to the ground state energy. We found a ground 
state energy of $ - 3/24 $ in (\ref{partads}), the ghosts contribute with
$ 2/24 $ and the internal CFT with $- c_{int}/24 = (c_{SL(2,R)} - 26)/24$. 
Using $c_{SL(2,R)} = 3 + {6 \over k-2}$, we find the 
 overall factor\footnote{Note that $c_{int}\geq 0$, $k>2$,  and
\ctotal\ imply that there will always be a tachyon in the theory.}
\eqn{..}{ 
(q\bar{q})^{-(1+c_{int})/24}
= e^{4\pi \tau_2 \left(1-{1 \over 4(k-2)}\right)}.
}

\begin{figure}[htb]
\begin{center}
\epsfxsize=5.0in\leavevmode\epsfbox{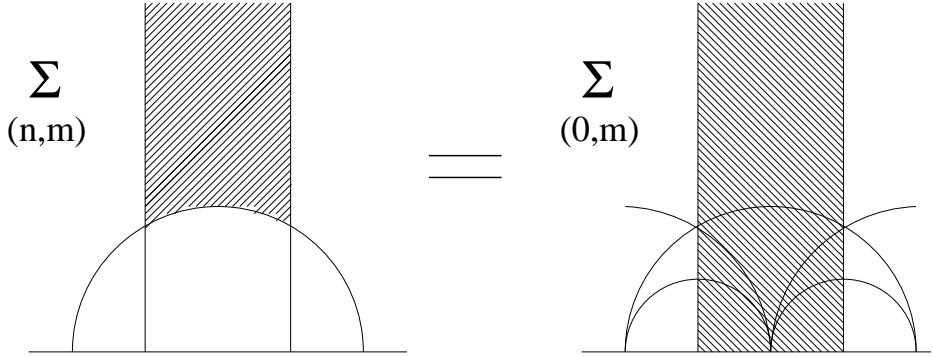}
\end{center}
\caption{The sum over $n$ is traded for the sum over copies of
the fundamental domain.}
\label{fig:fig0}
\end{figure} 

After multiplying (\ref{partads}) by the 
 $(b,c)$ ghosts  and the internal CFT partition functions, we should
integrate the resulting expression over the fundamental domain $F_0$ of the
modular parameter $\tau$.  The computation is much facilitated by
the trick invented in \cite{Polchinski}
to trade the sum over $n$ in (\ref{partads}) for the sum
over copies of the fundamental domain. See Figure 1. 
This is possible since $(n,m)$ transforms as a doublet
under the  modular group $SL(2,Z)$. 
If $(n,m)\neq (0,0)$, it can be mapped by an $SL(2,Z)$ 
transformation to $(0,m)$, $m>0$. The $SL(2,Z)$ transformation
also maps the fundamental domain into the strip ${\rm Im}\ \tau \geq 0$, 
$|{\rm Re}\ \tau| \leq 1/2 $.
On the other hand, $(n,m) = (0,0)$ is invariant under
the $SL(2,Z)$ transformation, and the corresponding term still has to be 
integrated over the fundamental domain $F_0$. This 
term represents the zero temperature contribution to the cosmological
constant, or the zero temperature vacuum energy. 
In addition to the usual tachyon 
divergence of bosonic string theory at large
$\tau_2$, it is also divergent due to the $\sin ^{-1}$ factor
in (\ref{partads}); this divergence can be interpreted as  coming from 
the infinite volume of $AdS_3$.  Since the temperature dependence 
of this term is  
trivial we will ignore it from now on.  
The final result then is that we fix $n=0$ in (\ref{partads}) 
and we integrate over the entire strip shown in Figure 1. 
%
%  However, this approach has a number of 
%difficulties.  The first is that as it stands in (\ref{partads}) there is 
%a complicated double sum, with $\tau$ dependence.  The second difficulty is
%that the integration region is not so simple.  As we will discuss, there 
%are poles in (\ref{partads}) that need to be regulated, and it is natural to
%divide the integration region into strips to do so.  This will be extremely
%difficult to deal with since $F_0$ has an awkward boundary.  Actually, these
%issues are not independent.  It is precisely the summation over 
%$(n,m)$, enforcing invariance under the full modular group
%\eqn{sl2z}{
%\left[ \begin{array}{c}
%n \\ m \end{array} \right] \rightarrow
%\left[ \begin{array}{c} 
%n'  \\ m'  \end{array} \right] 
%=\left[ \begin{array}{cc}
%a & b \\ c & d \end{array} \right]
%\left[ \begin{array}{c}
%n \\ m \end{array} \right], \ \ \ \ \ \ \ \left[ \begin{array}{cc}
%a & b \\ c & d \end{array} \right] \in SL(2,Z)
%}
%that restricts the range of $\tau$ from the entire upper half-plane to $F_0$.
%If we choose instead to sum over $(0,m)$, then only the transformations
%with
%\eqn{}{
%\left[ \begin{array}{cc}
%1 & 0 \\ 1 & 1 \end{array} \right]
%}
%corresponding to $\tau \rightarrow \tau +1$ survive, and we may 
%integrate over the larger region $|\tau_1|<{1\over 2}$, $\tau_2> 0$.  
The one-loop partition function of bosonic string 
theory on $H_3/Z \times {\cal M}$ is then
\ber  \label{part}
Z (\beta,\mu) &= & {\beta(k-2)^{{1\over 2}}\over 8 \pi}
 \int_0^\infty {d\tau_2 \over  \tau_2^{3/2}}
\int_{-1/2}^{1/2} d\tau_1  e^{4\pi \tau_2 
\left(1-{1 \over 4(k-2)} \right)} \sum_{h,\bar{h}} 
D(h, \bar{h})q^h \bar{q}^{\bar{h}}    \nonumber \\
  &  & \times \sum_{m=1}^{\infty} {e^{-(k-2)m^2 \beta^2 
/ 4 \pi \tau_2} \over |\sinh(m\hat{\beta} /2)|^2} 
\left| \prod_{n=1}^{\infty} {1-e^{2\pi i n \tau} \over 
(1-e^{m\hat{\beta} + 2\pi i n \tau})
(1-e^{-m\hat{\beta} +2\pi i n \tau})} \right| ^2.
\eer

\section{Reading off the spectrum}

We will now extract the spectrum of Lorentzian string theory on $AdS_3$ by
interpreting the one-loop partition function in the spacetime theory. 
The one-loop partition function is the single particle contribution to the 
spacetime thermal free energy, $Z(\beta,\mu) = - \beta F$.
To this each string state makes a contribution 
$\beta^{-1} \log(1-e^{-\beta (E + i \mu \ell) })$, 
where $E$ and $\ell$ are the energy and the angular
momentum of the state. The total free
energy is the sum over all such factors: 
\eqn{free}{
 F(\beta,\mu) = {1\over \beta} \sum_{string \in {\cal H}} \log 
\left(1-e^{-\beta
(E_{string} + i \mu \ell_{string}) }\right)  = 
 \sum_{m=1}^{\infty} f(m \beta, m\mu),}
where
\eqn{meq1} {
 f(\beta,\mu) = 
 { 1\over \beta } \sum_{string \in {\cal H}}
e^{-\beta( E_{string} + i \mu \ell_{string}) }.}
Here ${\cal H}$ is the physical Hilbert space of single
string states.
In both (\ref{part}) and (\ref{free}), we have the
sums over functions of $(m\beta, m\mu)$. It is therefore
sufficient to compare the $m=1$ terms in these expressions. 
In other words, we want
to verify that $E_{string}$ and $\ell_{string}$ extracted
from the identification, 
\ber \label{readof}
&& f(\beta,\mu) = \sum_{string \in {\cal H}}
 { 1\over \beta} e^{-\beta (E_{string} + i\mu \ell_{string}) } \nonumber \\ 
& = & {(k-2)^{{1\over 2}}\over 8\pi  } 
\int_0^\infty {d\tau_2\over \tau_2^{3/2}} 
\int_{-1/2}^{1/2} d\tau_1  e^{4\pi \tau_2 
\left(1-{1 \over 4(k-2)} \right)} \sum_{h,\bar{h}} 
D(h, \bar{h}) q^h \bar{q}^{\bar{h}} \nonumber \\
&&~~~~~~\times  {e^{-(k-2) \beta^2 / 4 \pi \tau_2} \over 
|\sinh(\hat{\beta}/ 2)|^2} 
\left| \prod_{n=1}^{\infty} {1-e^{2\pi i n \tau} \over
 (1-e^{ \hat{\beta} + 2\pi i n \tau})
(1-e^{- \hat{\beta} +2\pi i n \tau})} \right| ^2
\eer
agree with the string spectrum found in our previous paper
\cite{Maldacena:2000hw}.
We will see that the sum over the Hilbert space breaks up into a sum over the
discrete states and an integral over the continuous states, with the 
expressions for the energies that were reviewed in section 2.
Since the one-loop calculation presented here is independent of 
the assumptions made in \cite{Maldacena:2000hw} about strings 
in Lorentzian
$AdS_3$, we can regard this as a derivation of the spectrum starting 
from the well-defined Euclidean path integral. 

\subsection{Qualitative analysis}

In this subsection we will analyze (\ref{readof}) in a qualitative way and
explain where the different contributions to the spectrum come from. To 
keep the notation simple, we set $\mu=0$ in this subsection, leaving
the exact computation for the next subsection.

As expected, in (\ref{readof}) there is an exponential divergence as $\tau_2
\rightarrow \infty$, coming from the tachyon.  This is just as in the flat
space case, where $({\rm mass})^2 < 0$ of the tachyon causes its
contribution to be weighted with a positive exponential. We will disregard
this divergence\footnote{
A skeptical reader could think that we are really doing
the superstring partition function (the fermions included in 
the internal CFT, etc.).
Then  the tachyon divergence will disappear but one would still 
find the divergences that we discuss below. Of course, the one-loop
partition function is non-vanishing even in the supersymmetric case since
the thermal boundary conditions break supersymmetry.}.   
However, rather unexpectedly,  the
expression above has additional divergences at finite values of $\tau$.
In string theory one might naively expect that divergences come only
from the corners of the fundamental domain in the $\tau$-plane, 
but in this case the divergence is 
coming from points  in the interior of the fundamental domain. 
Overcoming the initial panic, one realizes that these divergences are
related to the presence of long strings. In fact, as with any other
string divergence, it can be interpreted as an IR effect. This divergence
is due to the fact that long strings feel a flat potential as they go to 
infinity and therefore we get an infinite volume factor. 
To see this, note that near the pole (see Figure 2)
\eqn{...}{
\tau= \tau_{pole} + \epsilon,}
where
\eqn{,,}{ 
\tau_{pole} = 
{r \over w} + i {\beta \over 2 \pi w} ,
}
we can expand the partition function and replace $\tau$ in  all terms by its
value at the pole, except in the one term  that has the pole.  
%Here $r$ is an integer such that
%$-{1 \over 2} \leq {r \over w} \leq {1 \over 2}$. 
\begin{figure}[htb]
\begin{center}
\epsfxsize=3.0in\leavevmode\epsfbox{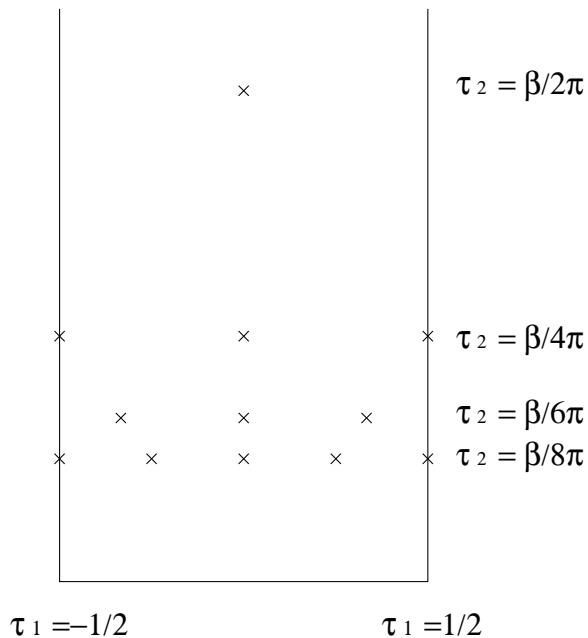}
\end{center}
\caption{Poles in the $\tau$ plane, shown for $w=1$ to 4.}
\label{fig:fig1}
\end{figure} 
If we integrate (\ref{readof}) near the pole, i.e. in the region
  $ \epsilon < | \tau - \tau_{pole}| \ll 1 $ , we find that it
diverges as $\log \epsilon$ with coefficient  
\eqn{longres}{
 {1 \over\sqrt{ w \beta^3}} 
\exp\left[ - \beta\left(  {k\over 2}w 
+{1 \over w} ( \tilde{N} + h + \tilde{\bar{N}} +\bar{h}-2 + {1\over 2(k-2)}  
)\right) 
+{2\pi i r \over w}(\tilde{N}+h-\tilde{\bar{N}}-\bar{h})\right].} 
We now sum over $r$, with $| r/w| \leq 1/2$, since these are the
ones corresponding to  the poles
in the strip\footnote{If some poles are on the boundaries of the strip,
  $\tau_1 = \pm 1/2$, then  we only
count them once. }. This sum constrains
$\tilde N+h- \tilde{\bar{N}}-\bar{h}$
  to be an integer  multiple of $w$, as in 
 \levelml , and  it introduces an additional factor
 of $w$ in \longres . The log divergence in
$\tau$-integral can therefore be expressed as
\eqn{contcontr}{ f(\beta,\mu)\sim  
  {1 \over \beta} \log \epsilon  \int_0^\infty ds  e^{- \beta E(s)}
+ \cdots ,} 
where $E(s)$ is the energy spectrum given by (\ref{econt}).
Note that the $s$-integral and the sum over $r$ we mentioned above 
 give the factor $\sqrt{w/\beta}$ needed
to match the prefactor
 in (\ref{longres}) to that 
in (\ref{contcontr}).   
This reproduces the expected contribution from the long strings 
in the left  hand side of
(\ref{readof}). The logarithmic divergence  should 
be interpreted as a volume factor due to the fact that the
long string can be at any radial position. In the next subsections, 
we will see more precisely that it is indeed associated to the 
infinite volume in spacetime by relating $\epsilon$ to a 
long distance cutoff.

Now we would like to calculate the short string spectrum.
Since the long string spectrum gives a divergent result, while the 
short string spectrum gives a finite one, it might appear at first that 
extracting the contributions due to the short strings from a divergent
expression such as (\ref{readof}) will be problematic.
Fortunately we can get around this difficulty since the
temperature dependence of the long string free energy is different
from that of the short string free energy.
In the next subsection we will explain how to do this precisely and 
reproduce the short string spectrum which agrees with 
\cite{Maldacena:2000hw}. One of the more
puzzling aspects of the short string spectrum found there is
that there is a cutoff $1/2 < \tilde{j} < (k-1)/2$
in the value of the $SL(2,R)$ spin $\tilde j$.
 In the remainder of this section we will explain in 
a qualitative way how this cutoff arises by doing the calculation for
large $k$.
 
If we were to evaluate the right hand side of (\ref{readof}) 
 naively (and incorrectly),
 we would expand the integrand in  
powers of $q=e^{2\pi i \tau}$ and then perform the $\tau$ integral. 
If we did this, we would obtain the short string spectrum with $w=0$ and 
no upper bound on the value of $\tilde j$. However this expansion 
is not correct.
How we can expand the integrand in (\ref{readof}) depends on the 
value of $\tau_2$. When we 
cross the poles at $\tau_2 = {\beta \over 2\pi w}$, a different expansion
 should be used for the denominator:
\ber
    {1 \over 1 - e^{\beta+2\pi i w \tau}}
 &=&~~ \sum_{q=0}^\infty e^{q(\beta + 2\pi iw \tau)},~~~~~~~~~
\left( \tau_2 > {\beta \over 2\pi w} \right), \cr
&=& -\sum_{q=0}^\infty e^{-(q+1)(\beta + 2\pi iw \tau)},~~~
\left( \tau_2 < {\beta \over 2\pi w} \right).\eer
 When $\tau_2$ is in the range
\eqn{taurange}{
{\beta \over 2\pi (w+1)} < \tau_2 <{\beta \over 2\pi w}, 
}
the product over $n$ in the first term in the denominator in 
 (\ref{readof})  is broken
up into two factors, a product in $1 \leq n \leq w$ and a product
in $w+1 \leq n$. The first factor is expanded in powers of
$e^{-(\beta + 2\pi i n\tau)}$ and the second factor
is expanded in powers of $e^{\beta + 2\pi i n \tau}$. Combining
them together with the terms coming from the expansion 
of the remaining products in (\ref{readof}),
we get an exponent of the form\footnote{
The first term $-\beta/2$ comes from expanding
$1/\sinh(\beta/2)$ in (\ref{readof}). }
%\ber
%& &-{1 \over 2} \beta - \sum_{n=1}^w (q_n + 1)(\beta + 2\pi i n \tau)
%  + \sum_{n=w+1}^\infty q_n (\beta + 2\pi i n \tau) \nonumber \\
%& =& - \left({1\over 2} + q + w  \right)
%\beta + 2\pi i\tau \left(N_w -{1 \over 2}w(w+1)\right),
%\eer
%where\footnote{The first term $-\beta/2$ comes from
%$1/\sinh(\beta/2)$ in (\ref{readof}). }
%\eqn{.,.,}
%{ q = \sum_{n=1}^w q_n - \sum_{n=w+1}^\infty q_n,
%~~~N_w = -\sum_{n=1}^w n q_n + \sum_{n=w+1}^\infty n q_n.} 
\eqn{expoform}{- \left({1\over 2} + q + w  \right)
\beta + 2\pi i\tau \left(N_w -{1 \over 2}w(w+1)\right),
}
for some integers $q$ and $N_w$. 
There is a similar term for $\tau \rightarrow \bar{\tau}$.
We are then to do the $\tau$-integral of the form,
\eqn{...,,,}{
 \int {d^2\tau \over \tau_2^{3/2}}
   e^{
  4\pi \tau_2 \left(1 - {1\over 4(k-2)}\right) - (k-2){\beta^2 \over
4\pi \tau_2} -  \beta (1+  q + \bar{q} + 2 w ) + 
2\pi i \tau \left(N_w + h -{1 \over 2}w(w+1)\right)
- 2\pi i\bar{\tau}\left(\bar{N}_w + \bar{h} 
- {1 \over 2}w(w+1)\right)},}
over the region (\ref{taurange}). 
The integral over $\tau_1$ produces the level matching condition \levelms .
Now we evaluate the integral over $\tau_2$ using the saddle point
approximation.  We find that the saddle point is at
\eqn{saddle}{
\tau_{saddle}={(k-2) \beta \over 
2\pi\sqrt{1+4(k-2)(N_w +h-1-{1\over 2}w(w+1) ) } }
}
and the integral gives
\eqn{....}{
{1 \over \beta}\exp\left[- \beta \left(1+q+\bar{q}+2w
+\sqrt{1+4(k-2)(N_w+h-1-{1\over 2}w(w+1))} \right)\right].
}
This is the correct form of the contributions due to
the short strings in the left hand side of (\ref{readof}).  
Moreover we obtain the bound on $\tilde{j}$ precisely,
because $\tau_{saddle}$ has
to be in the range (\ref{taurange}) in order for the
saddle point approximation to give a non-zero result. 
By (\ref{saddle}), this condition is the same as
the bound on the spectrum (\ref{rangedisc}), which
is equivalent to  $1/2 < \tilde{j} < (k-1)/2$.
(It is a bit surprising that we get all factors precisely right from
the saddle point approximation.)
%
%\eqn{cond}{
%{k-2\over 2}(w+1)  >\sqrt{{1 \over 4}+(k-2) 
%\left(N+h-1-{w(w+1) \over 2} \right) } > {k-2\over 2}w.
%}
%Using the mass shell condition $L_0=1$, this implies 
%${k-1 \over2} >j> {1\over2}$. 
Notice then that the cutoff in $\tilde{j}$ 
is associated to the fact that 
we expand the integrand in (\ref{readof}) in different ways depending 
on the value of $\tau$. The value of $\tau $ making 
the biggest contribution 
to the integral depends on the values of $N$ and $h$ of the string state.

\subsection{A precise evaluation of the $\tau$-integral}

Now let us study the partition function (\ref{readof}) more 
systematically.  
In this subsection, we go back to the general case with 
$\mu\neq 0$. From what we saw in the previous subsection, 
we expect to find the discrete
states from the integral over the range (\ref{taurange}),
and the continuous states from the poles
after a suitable regularization. 
% To simplify the expressions we will find 
% it convenient to shift $h$ and $\bar{h}$ by 1.
 
%The $SL(2,R)$ character may be expanded as
%\eqn{}{
%\sum_{N, J^3} q^N e^{m\beta J^3}
%}
%with $-N \leq J^3 \leq N$. 
%In analogy with (\ref{free}) we denote by $Z_m(\beta)$ the term in the 
%expansion of the partition function that has an overall coefficient 
%$(m\beta)^{-1}$.
In order to evaluate the $\tau$-integral exactly, it is useful
to introduce a new variable $c$ by
\eqn{.....}{
e^{-(k-2){ \beta^2 \over 4 \pi \tau_2}} = -{8\pi i \over \beta} 
\left( {\tau_2 \over k-2} \right)^{{3 \over 2}} \int_{-\infty}^{\infty} 
dc \ c \  e^{-{4\pi \tau_2 \over k-2} c^2 + 2i \beta c} \ .
}
Now suppose $\tau_2$ is in the range,
\eqn{taurangen}{
{\beta \over 2\pi (w+1)} < \tau_2 <{\beta \over 2\pi w} ,
}
and expand the integrand in (\ref{readof}) as explained in the
 previous subsection.
% The integrand has the form
%\eqn{integrand}{
% e^{ - \beta ( 1/2 +  w + q ) - \bar \beta (1/2 +  w + \bar q  }
% e^{ 2 \pi i \tau ( N_w + h -1 -
%w(w+1)/2 + { 1 \over 4 (k-2)} ) } e^{ 2 \pi i \bar \tau ( \bar N_w + \bar h -1 -
%w(w+1)/2 + { 1 \over 4 (k-2)} ) }   
The right hand side of (\ref{readof}) becomes a sum of terms like 
%\eqn{}{
%Z(\beta) =  
%\sum D(h, \bar{h}, N, \bar{N}, J^3, \bar{J}^3, q, \bar{q}, w) Z_m(\beta),
%}
%with  
\ber \label{partm}
 & &  {4   \over  \beta (k-2) i} 
\int_{-\infty}^{\infty} dc \ c \
\int_{{ {\beta} \over 2\pi(w+1)} }^{{{\beta} \over 2\pi w} } 
d\tau_2 \int_{-1/2}^{1/2} d\tau_1  \nonumber \\
&& \times
 \exp\left[ - \hat{\beta} \left(q +  w + {1\over 2} \right) - 
\bar{\hat{\beta}} \left(\bar{q}+ w + {1\over 2} \right)
+ 2\pi i\tau_1 ( N_w + h - \bar N_w - \bar h )  \right.\nonumber \\
& &  ~~~~~~ \left.
 + 2ic \beta 
-2\pi \tau_2\left(h+ \bar{h}+ N_w + \bar{N}_w +{2c^2 + {1 \over 2}\over k-2}
 -w(w+1) -2\right)\right].
\eer
The integral over $\tau_1$ gives a delta function enforcing $N_w + h  = 
 \bar{N}_w + \bar{h} $, which is the level-matching condition 
 \levelms . 
Integrating over $\tau_2$ in the range \taurangen\  gives
\ber
& & 
{1 \over \beta\pi i}
\int_{-\infty}^{\infty} dc ~c~{
\exp\left[ 2ic \beta - \hat{\beta} \left( q +  w + {1\over 2}
\right) - \bar{ \hat{\beta} } 
\left( \bar{q} + w + {1 \over 2} \right)  \right]
 \over c^2 +{1 \over 4}+{(k-2)}\left( N_w + h -1  - {1\over 2}
w(w+1) \right)}
 \nonumber \\
& &~~~~ \times\left\{ - \exp\left[-{ \beta\over w} \left(  
2 N_w   + 2 h -2 +
{2c^2+ {1\over 2} \over k-2} -w(w+1)  \right) \right]
 \right. \nonumber \\
&&~~~~~\left. + \exp\left[ -{\beta \over w+1}
\left( 2 N_w + 2 h  -2  +{2c^2 + {1\over 2}\over k-2}  -w(w+1)  \right)
\right] \right\}
\label{intofc} \eer
where we used \levelms .

Let us first look at the first term  (the second line)
 in (\ref{intofc}). 
We note that the exponent can be expressed in the form
of a complete square if we set $c=s+{i\over 2}(k-2)w$.
As it will become clear shortly, 
it is natural to shift the contour of the $c$-integral
from ${\rm Im}~c=0$ to ${\rm Im}~ c = {1\over 2}(k-2)w$
so that $s$ becomes real. 
During this process the contour crosses some poles 
in the integrand,  picking up the residues of the poles 
in the range $0<{\rm Im} \ c \ < {1\over 2} (k-2)w$.  See Figure 3.
The poles are located at
\eqn{1residue}{
-{ c^2 \over (k-2) } = 
 N_w  + h  -{1\over 2}w(w+1)  -1+{1\over 4(k-2)} < {k-2 \over 4}w^2.
}
Similarly, for the second exponential term (the third line)
in (\ref{intofc}) we shift 
the contour to $c=s+{i\over 2}(k-2)(w+1)$ with $s$ real.
This picks up the poles at 
\eqn{2residue}{
- { c^2 \over (k-2) } =  N_w + h - {1\over 2}
w(w+1) -1 +{1\over 4(k-2)}  < 
{k-2 \over 4}(w+1)^2.
}
It is important to note that the residues of these poles 
have a sign opposite to that of the
residues of the poles obeying (\ref{1residue}).
So the result is that we are left with only those poles in the range 
\eqn{boundonc}{
      {k-2\over 2} w ~ < ~{ \rm Im} ~c~  < ~{k-2\over 2}(w+1),
}
with residues
\eqn{precisedisc}{
{1 \over \beta} 
\exp\left[-\hat{\beta} q - \bar{ \hat{\beta}} \bar q - \beta 
 \left(1  +2w
+\sqrt{1+4(k-2)(N_w + h -1-{1 \over 2}w(w+1))}\right)\right].
}
This is the expected contribution of the short strings to the right 
hand side of (\ref{readof}), and we see also that \boundonc\   
translates into the correct bound on $\tilde{j}$ \rangej .
\begin{figure}[htb]
\begin{center}
\epsfxsize=4.0in\leavevmode\epsfbox{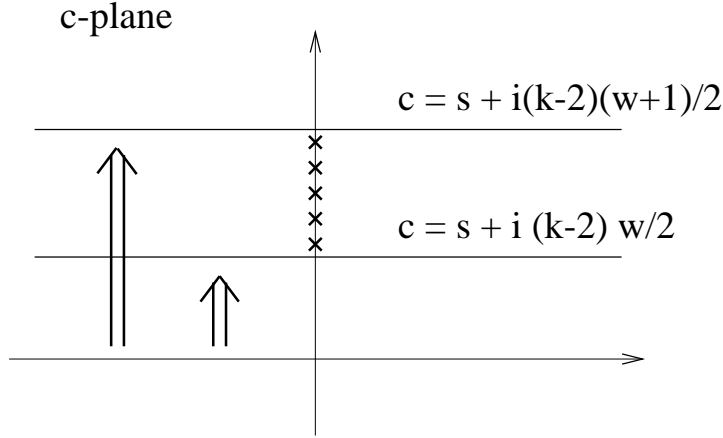}
\end{center}
\caption{Shifting the contour of integration picks up the pole
residues corresponding to the short string spectrum.}
\label{fig:fig2}
\end{figure}

It remains to examine the resulting integral over $s$.  
Notice that the term coming from just above the pole at 
$\tau = \hat{\beta}/ 2 \pi w$ has a very similar
 $w$ dependence in the exponent as that coming from just below the 
pole. In other words, we combine the first term of (\ref{intofc}) with 
the second term of an expression similar to (\ref{intofc})
 but with $w \to w-1$
and we find, after shifting the countours as above, 
\ber \label{wlong}
&& {1 \over 2\pi i\beta}  \int_{-\infty}^{\infty} 
ds  \left({2s \over w(k-2)}+ i \right) \nonumber \\
&&~~~\times \left(
{\exp\left[- \hat{\beta} q - \bar{\hat{\beta}} \bar q 
- \beta \left( {k \over 2}w + {2\over w }
\left({s^2 + 1/4 \over k-2} + N_{w-1} + h -1
\right)\right)\right] 
\over  {1 \over 2} +is-  {k\over 4}w+{1 \over w}\left(N_{w-1}+ h -1
 + {s^2 +1/4 \over k-2}\right)}  \right. \nonumber\\
&&\left. ~~~~~~~~~~
 - {\exp\left[- \hat{\beta} q - \bar{ \hat{\beta}}
 \bar q - \beta \left( {k \over 2 }w 
  + {2\over w }
\left({s^2 + 1/4 \over k-2} + N_{w} + h -1 \right)\right) \right]
 \over  -{1 \over 2} +is-  {k\over 4}w
+{1 \over w}\left( N_w + h -1 
+  {s^2 +1/4 \over k-2}\right)} \right) .
\eer
Let us concentrate for now on the third line of (\ref{wlong}).
We first note that the sum of such terms over all states
gives rise to the log divergence. To see this,
it is useful to notice that the combinations 
\eqn{comb}{ \tilde{N} =
 q w + N_{w}  ~,~~~~~~~~~ \tilde{\bar N} = \bar q w+ \bar N_{w} }
that appear in the exponent of the third line of equation
%\footnote{
%Here we are using the level-matching condition $N_w + h =
%\bar{N}_w + \bar{h}$.} 
(\ref{wlong})
are the levels before spectral flow. 
Thus, for a given state $| \psi \rangle$,
states of the form  $( \tilde J_0^+ \tilde{{\bar J}_0^+}
)^n |\psi\rangle $ all have
the same value of $ \tilde N$ and  $\tilde {\bar  N} $.
Acting with $\tilde J_0^+ \tilde{{\bar J}_0^+}$ on 
$|\psi \rangle$
does not change  the exponent in (\ref{wlong}), but it
does change the denominator by one.  This implies that when 
we sum over all the states of this type,
we will find a divergent sum of the form
$$\sum_{n=0}^\infty { 1 \over A - n }.$$ 
This divergence has the same origin as the divergence  of the
right hand side of (\ref{readof}) at the pole $\tau_{pole} =  \hat{\beta}
/ 2 \pi w$. In fact, if we regularize the $\tau$-integral by removing
a small region near the pole as $|\tau - \tau_{pole}| > \epsilon$,
we find an additional factor $e^{ - n \epsilon}$ in 
the sum. In the next subsection, we will give the spacetime 
interpretation of this procedure. With this regularization,
the sum behaves as $ \log \epsilon $. More precisely we have 
\eqn{divlong}{
- \sum_{n=0}^\infty { 1 \over A -n } e^{- n \epsilon} =
 \log \epsilon  + { d \over dA } \log \Gamma(-A)  + { \cal O}( \epsilon)
}
where 
\eqn{defa}{
A = - { 1 \over 2} + i s - { k \over 4 }w + {1 \over w} 
\left(  { s^2 + {1\over 4} \over k-2}+ \tilde N +  h -1 \right).
}    
Here we have assumed that 
\eqn{assumption}{
  \tilde {\bar N} + \bar h 
\leq \tilde N +  h, 
}
but it can be seen that the other case gives the same result. 

Now we turn our attention to the second  line of (\ref{wlong}).
In those terms we have one less unit of spectral flow, as compared to 
the third line in (\ref{wlong}) that we analyzed above. 
In other words, now we will have that
$ (w-1) q + N_{w-1} = \tilde {N'} $.
These states are in the spectral flow image of ${\cal D}_j^+$. 
Since we want to combine these states  with the states  coming from the 
third line in (\ref{wlong}) it is convenient to do spectral
flow one more time and think of these states as in the spectral flow
image of ${\cal D}_j^-$ under $w$ units of spectral flow. In this 
case we find that $ q + \tilde {N'} = \tilde N $ where now 
$ \tilde N $ is the level of the ${\cal D}_j^-$ representation before
spectral flow. From now on the discussion is very similar to what we
had above. The states with $ ({\tilde J}_0^- \tilde{{\bar J}_0^-} )^n 
|\psi \rangle $ all have the same energies but they will contribute to the
denominator of the second line in (\ref{wlong}) with 
\eqn{densecond}{
\sum_{n=0}^\infty { 1 \over B + n}  e^{- n \epsilon} 
 =  \log \epsilon  - { d \over d B} \log \Gamma(B) + {\cal O}(\epsilon)
}
where 
\eqn{defb}{
B =  { 1 \over 2} + i s -  { k \over 4 }w + {1 \over w} 
\left(  { s^2 + {1 \over 4} \over k-2}+\tilde{ \bar N} +  \bar h -1 
\right),
}
again assuming \assumption .
%
%Let us sum over the modes $J^3=\alpha$ , $N=\alpha (w+1)$, 
%which keeps the exponential fixed.  Using the formula
%\eqn{}{
%\sum_{\alpha=0}^N {1\over a+\alpha}= \log N - {d \over da} \log \Gamma (a),
%}
%(\ref{wlong}) can be written as 

%with $E(s)$ the energy of long strings (\ref{econt}) and the 
%density of states $\rho(s)$  given by
After we perform these two sums, we find that (\ref{wlong}) can be written 
in the form 
\eqn{precisecont}{
{2 \over \beta } \int_0^{\infty} ds \rho(s) \exp\left[ -  \beta\left(  
E(s) + i {\mu \over w}(\tilde N + h - \tilde {\bar N} - \bar h) \right)
\right] 
}
with $E(s)$ the energy of long strings (\ref{econt}) and 
$\rho(s)$ the  density of states.
We will later  see that the physical momentum $p$ of a long string in 
the $\rho$ direction is equal to $p = 2 s$.
The angular momentum $\ell =(\tilde N + h - \tilde {\bar N} - \bar h)/w$
is an  integer since  the  states in (\ref{wlong}) 
were obeying \levelms\ and the definition \comb\ ensures that 
\levelml\ is satisfied. 
The density of states  $\rho(s)$ derived from this analysis is
\eqn{density}{
\rho(s) = {1 \over 2 \pi }  2 \log \epsilon + { 1\over 2 \pi i}
 {d \over 2d s } \log \left( 
{\Gamma({1\over 2}-is+ \tilde{\bar m}) \Gamma({1\over 2}-is- \tilde m) 
\over
\Gamma({1\over 2}+is+\tilde {\bar m} ) \Gamma({1\over 2}+ is-\tilde {m})}
\right) ,
}
where
\eqn{.......}{
\tilde m=-{k \over 4}w +{1\over w} \left( {s^2+{1\over4} \over k-2} + 
\tilde N +h -1 \right) ,~~~~
\tilde{\bar m} =-{k \over 4}w+{1\over w} \left( {s^2+{1\over4}
 \over k-2} + \tilde {\bar N} + \bar h -1 \right) .
}
Note that, despite appearances to the contrary, \density\ is 
actually symmetric under $\tilde m \leftrightarrow \tilde{\bar m}$ 
since $\tilde m - \tilde {\bar  m} = \ell$ is an integer.
In the next subsection we will show 
that this density of states  \density\ 
is what is expected from the spacetime 
meaning of the cutoff $\epsilon$. In going from 
(\ref{wlong}) to \precisecont\ we
have states which could be interpreted as coming from the spectral flow
of the discrete representations ${\cal D}_j^+ $ and
${\cal D}_j^-$, with the zero modes
essentially stripped off since they were explicitly summed over
in \divlong\ and \densecond . This implies that the states we have
in the end belong to the continuous representation. Note also
that the integral over $s$ in \precisecont\ has only half the range in 
(\ref{wlong}). We rewrote it in this way using the fact that the exponent
is invariant under $s \to -s$, and that is the reason why we have four
Gamma functions in \density .
% In other words, if we have 
%$\int_{-\infty}^\infty ds h(s) e^{-Bs^2}$ in (\ref{wlong}) then we write 
%$\rho(s) = h(s) + h(-s)$ in \precisecont . 
In going from (\ref{wlong}) to \precisecont\
we have also used that $ {d \over d A} = {1 \over {d A(s) \over ds}}{d
\over ds }$ in \defa\ and similarly in \defb .

Combining eqns. (\ref{precisedisc}) and (\ref{precisecont}), we have finally
\eqn{........}{
f(\beta, \mu)= {1\over \beta } \sum 
D(h, \bar{h}, \tilde N, {\tilde {\bar{N}}}, w)
\left[\sum_{q, \bar{q}}e^{- \beta( E +i \mu\ell) } 
+ \int_0^\infty ds \rho(s)e^{- \beta( E(s) +i \mu \ell)  } \right]
}
which is the free energy due to the short strings and the long strings, 
respectively.

\subsection{The density of long string states}

What remains to be shown is the interpretation of  $\rho(s)$
given by (\ref{density}) as  the density of long string
states. Whenever we have a continuous spectrum the density of states may be
calculated by first introducing a long distance cutoff which will make the 
spectrum discrete, and then removing the cutoff. If the cutoff is related 
to the volume of the system then the density of states will have a 
divergent part, proportional to the volume and dependent only on the bulk
physics, and a finite part which encodes information about the scattering
phase shift and also has some dependence on the precise cutoff procedure. 
% In a quantum mechanical system,
%a density of scattering states is related to a phase shift. 
To see this, let us consider a one-dimensional
quantum mechanical model on the half line, $\rho > 0$, with a potential
$V(\rho)$. We assume that $V(\rho)$ vanishes sufficiently fast 
for large $\rho$, and that there is continuous spectrum above a certain
energy level. To define the density of states, it is convenient 
to introduce a long distance cutoff at large $\rho$ so 
that the spectrum becomes discrete. 
Let us first consider a cutoff by an infinite wall at $\rho=L$.
If $L$ is sufficiently large,  an energy eigenfunction $\psi(\rho)$
near the wall can be approximated by the plane wave 
\eqn{,,,}{ \psi(\rho) \sim e^{-i p \rho} 
+ e^{ip\rho +i\delta(p) },
}
where $\delta(p)$ is the phase shift
due to the original potential $V(\rho)$.  Imposing 
Dirichlet boundary condition $\psi(L) = 0$ at the wall, we have 
\eqn{wall}{
    2pL + \delta(p) = 2\pi \left(n + {1\over 2}\right)}
for some integer $n$. 
If $L$ is sufficiently large, there is a unique solution $p=p(n)$
to this equation for a given $n$. As we remove the
cutoff by sending $L \rightarrow \infty$, the spectrum of $p$ becomes
continuous. We then define the density of states $\rho(p)$ by
\eqn{dos}{
   dn =  \rho(p) dp.} From (\ref{wall}), we obtain
\eqn{dosmore}
{  \rho(p) = {1 \over 2\pi}\left( 2L + {d\delta \over dp} \right). }
Thus the finite part of the density of states is given by
the derivative of the phase shift. 

Instead of the infinite wall at $\rho =L$, we may consider
a more general potential $V_{wall}(\rho-L)$ which vanishes for
$\rho < L$ but rises steeply for $L < \rho$ to confine the particle. 
Let us denote by $\delta_{wall}(p)$ the phase shift due to scattering from 
$V_{wall}(\rho)$. 
We then obtain the condition on the allowed values of momenta by 
matching these two wavefunctions and their derivatives at  $\rho=L$ as
\eqn{,.,.,.}{
\psi(\rho) \sim  e^{-ip\rho} + e^{ip\rho + i\delta(p)} \sim A \left[
 e^{-ip(\rho-L)} + e^{ip(\rho-L)+i\delta_{wall}(p)} \right], ~~~(\rho \sim L).}
It follows that
\eqn{match}
{ p L + \delta(p) = -pL + \delta_{wall}(p) + 2\pi n.}
In the limit $L\rightarrow \infty$, the density of states given
by $dn = \rho(p) dp$ is then
\eqn{doswall}{
  \rho(p) = {1 \over 2\pi} \left( 2L +{d\delta \over dp}
              - {d\delta_{wall} \over dp} \right).}
When we have the infinite wall, the phase shift due to the wall
is independent of $p$ ($\delta_{wall} = \pi$), 
and (\ref{doswall}) reduces to (\ref{dosmore}). 
  
In order to apply this observation to our problem, it is useful 
to first identify the origin of the logarithmic divergence in the one-loop 
amplitude $Z (\beta,\mu)$ by examining the functional 
integral (\ref{path}) near the boundary
of $AdS_3$. In the cylindrical coordinates (\ref{cyl}),  
the string worldsheet action (\ref{act}) for large $\rho$
takes the form
\eqn{actcyl}
 { S \sim {k \over \pi} \int d^2 z \left( \partial \rho \bar{\partial}\rho
+ {1 \over 4} e^{2\rho}  |\bar{\partial} (\theta-it)|^2
+ \cdots \right).}
Because of the factor $e^{2\rho}$, the functional
integral for large $\rho$ restricts  
$(t, \theta)$ to be a harmonic map from the worldsheet
to the target space. Since $(t,\theta)$ are coordinates 
on the torus,   
\eqn{,,,,}{ \theta -it \sim  \theta - i t + 2\pi n + i \hat{\beta} m,
~~~(n, m \ {\rm integers}),}
the harmonic map from the torus to the torus is 
\ber\label{harmonic}
   \theta - it
&=&  ( 2\pi w + i \hat{\beta}  m )
    \sigma^1 + 
 (2\pi r + i \hat{\beta} n ) \sigma^2 \nonumber \\
&=&
\left[ (2\pi w + i \hat{\beta}  m ) \tau - 
 ( 2\pi r + i \hat{\beta}  n ) \right]
{\bar{z} \over 2 i \tau_2} \nonumber \\
&&~~~~~~~~~ -
\left[ ( 2\pi w + i \hat{\beta} m ) \bar{\tau} - 
 ( 2\pi r + i \hat{\beta} n )\right]
{z \over 2 i \tau_2},
\eer
where $z = \sigma^1 + \tau\sigma^2$ is the worldsheet
coordinate and $(r,w,n,m)$ are integers. 
In particular, the map  
 $(\theta - it)$ with $(n,m)=(1,0)$ 
becomes $w$-to-$1$ and {\it holomorphic}
when $\tau$ takes the special value 
\eqn{,,,,,}{ \tau_{pole} = {r \over w} + i 
{\hat{\beta} \over 2\pi w} .}
On the other hand, if $\tau$ is not at one of these
points, $\bar{\partial}(\theta-it)$ cannot be set to 
zero\footnote{
For any $\tau$, we also have a trivial holomorphic map
$(t,\theta)={\rm const}$. The functional integral around
such a map gives a result independent of ${\beta}$
and we can neglect it in the following discussion.}.  
This gives rise to an effective potential $e^{2\rho}$ for $\rho$, 
which keeps the worldsheet from growing towards the boundary. 
If $\tau$ is near $\tau_{pole}$
\eqn{nearspecial}
   {\tau = \tau_{pole} + \epsilon,
}
the harmonic map (\ref{harmonic}) with $(n,m)=(1,0)$
gives
\eqn{,.,.}{
  |\bar{\partial}(\theta - i t)|^2 \sim 
\left( {2 \pi^2 w^2 \over \beta}\right)^2 \epsilon^2. } 
Thus the action (\ref{actcyl}) generates the Liouville potential
$\epsilon^2 e^{2\rho}$. When we computed the one-loop
amplitude in sections 4.1 and 4.2, we regularized the $\tau$-integral 
by removing a small disk $|\tau -\tau_{pole}| < \epsilon$ 
around each of these special points. Near $\tau = \tau_{pole}$,
this is equivalent to adding the infinitesimal Liouville 
potential $\epsilon^2 e^{2\rho}$ to the worldsheet action. For
$|\tau - \tau_{pole}| \gg \epsilon$, the worldsheet can never
grow large enough and the effect of the Liouville term is
negligible. To be precise, the Gaussian functional 
integral of $(t,\theta)$ shifts $k \rightarrow (k-2)$ as
in (\ref{poly}) and the effective action for $\rho$ near
$\tau = \tau_{pole}$ is 
\ber \label{effectiveliouville}
  S_{Liouville} = {k-2 \over \pi} \int d^2 z \left( 
\partial \rho \bar{\partial} \rho + \epsilon^2 e^{2\rho} \right).
\eer
Therefore, we find that our choice of regularization 
in \divlong\ and \densecond\ amounts to introducing the Liouville
wall which prevents the longs strings from going to very 
large values of $\rho$. By looking at the potential in 
(\ref{effectiveliouville}), we 
see that the effective length of the interval is $ L \sim 
\log \epsilon $.
The central charge of this Liouville theory is such that the $e^{2 \rho}$ 
term has conformal weight one, 
\eqn{cenliou}{
 c_{Liouville} = 1 + 6 \left( b + {1\over b}\right)^2 
~,~~~~~~~~ b \equiv { 1 \over \sqrt{k-2}} ~ .
 } 
% We also expect that the finite part of the 
%density of states should be 
%\eqn{finite}{
%\rho_{\rm finite}(s) = { d \over ds }( \delta(s) - \delta_L(s) )
%}
The finite part of the density of states will be given through 
(\ref{doswall}) by  $\delta(s)$, the phase shift in the $SL(2,R)$ model,
and $\delta_{wall}(s)$, the corresponding quantity in Liouville theory. 
The first one was calculated in \cite{Teschner:1997ft,zamnotes},
\eqn{sltwoph}{
 i \delta(s) = \log \left( 
{\Gamma({1\over 2}+is- \tilde m) \Gamma({1\over 2}+is+ \tilde {\bar m})
 \Gamma(-2is)
 \Gamma({2is \over k-2})
\over
\Gamma({1\over 2}-is- \tilde m) \Gamma({1\over 2}-is+ \tilde{ \bar m})
 \Gamma(2is) 
\Gamma({-2is \over k-2})}
\right),
}
while the second one was obtained in 
\cite{Dorn:1994xn,Zamolodchikov:1996aa}\footnote{
In order to compare with the expressions in \cite{Dorn:1994xn,
Zamolodchikov:1996aa} we use the value of 
$b$ given in \cenliou\ and note that the relevant values of 
$\alpha$ are $\alpha = Q/2 + i s b$.}
\eqn{liouph}{
 i \delta_{wall}(s) = \log \left( {\Gamma(-2is) \Gamma({2is \over k-2}) \over
\Gamma(2is) \Gamma({-2is \over k-2})} \right).
}
Using these two formulas we can check that indeed the  density of states
\density\  is given by (\ref{doswall}). 
We can view this as an independent calculation of \sltwoph\ or
as an overall consistency check. 
Notice that the physical momentum $p$ of a long string along the 
$\rho$ direction is $p=2s$. This can be seen by comparing 
the energy of a long string \econt\  with the energy expected from 
(\ref{effectiveliouville}) 
with spacetime momentum $p$ along the radial direction,
$p = (k-2) w \dot \rho $.  We have chosen the variable 
 $s$ since it is conventional to
denote by $j = 1/2 + i s $ the $SL(2,R)$ spin of a continuous representation.

\section*{Acknowledgements}

H.O.\ would like to thank J. Schwarz and 
the theory group at Caltech for the kind hospitality
while this work was carried out. 

The research of J.M.\ 
was supported in part by DOE grant DE-FGO2-91ER40654,
 NSF grant PHY-9513835, the Sloan Foundation and the 
 David and Lucile Packard Foundations. 
The research of H.O.\ was supported in part by 
NSF grant PHY-95-14797, DOE grant DE-AC03-76SF00098,
and the Caltech Discovery Fund. 

%\newpage

%\appendix

%\section{???}

\renewcommand{\baselinestretch}{0.87}
%\footnotesize

\bibliography{papertwo}
\bibliographystyle{ssg}

%\nocite{*}    %This command makes every ref in review.bib appear.

\end{document}